# Scripting data acquisition operations and choice of data format for the data files of the DUCK ultra-high energy cosmic rays detector

*Dmitriy* Beznosko[1*], *Farid* Gasratov[2], *Fernando* Guadarrama[3], and *Alexander* Iakovlev[3]

[1] School of Sciences, Clayton State University, Morrow, GA 30260 USA
[2] P. N. Lebedev Physical Institute of the Russian Academy of Sciences, Moscow, 119991 Russia
[3] College of CIMS, Computer Science department, Clayton State University, Morrow, GA 30260 USA
[4] Upper School Science Department, Woodward Academy, College Park, GA 30337 USA

**Abstract.** This document outlines the control software considerations for the D.U.C.K (Detection of Unusual Cosmic casKades). The primary goal of this software is to provide users with the ability to control Flash Analog to Digital Converter functions and conduct DAQ (Data Acquisition) operations as well as set the file format for saving the data. The ROOT software framework was found to be particularly useful for DAQ and serves as the primary tool for storing and analyzing our data. Limitations of the software are being considered, and further development is being conducted.

## 1 Introduction

Ultra-high-energy Cosmic Ray (CR) research has gained considerable interest in recent years. CR research holds great promise for advancing the field of particle physics and enhancing our understanding, particularly in uncovering the behavior, origin, and interactions of charged particles within our galaxy. The DUCK (Detector system of Unusual Cosmic ray casKades) system [1, 2] aims to detect cosmic rays with a high degree of certainty. Its ultimate purpose is to gather data and contribute to the methodology of EAS (Extensive Air Shower) event analysis. The research also seeks to provide independent confirmation for the detection of "unusual" CR events.

The Horizon-T detector system [3, 4] is located at the campus of Tien Shan High-Altitude Scientific Station of the Lebedev Physical Institute. It is located just south of the city of Almaty in Republic of Kazakhstan at an altitude of 3340 m above sea level. The goal of this experiment is to study the temporal structure of EAS that are caused by cosmic rays with primary particle energy above $10^{16}$ eV.

The aim of this paper is to provide software considerations for DUCK and Horizon-T. More specifically, it aims to give an overview of the specialized DAQ software design and delve into the DAQ aspects of the software.

## 2 Software Considerations

The design of the Data Acquisition software is primarily based on the hardware design of the prototype and the considerations for the final experimental setup [5, 6] as well as the design of the side experiments that were used to test the early versions of the software [7, 8]. The DAQ is designed to work with Flash Analog to Digital Converser (FADC) model DT5730 by CAEN [9]. Up to three FADCs are supported at the same time.

The software architecture was designed to prioritize high-speed readout, control all FADC functions, and provide a real-time display for viewing cosmic events, with priority given to data retrieval and saving.

The software employs the Intel TBB (Threading Building Blocks) [10] library to manage the data queue. A single thread monitors the FADC, checking for new data and retrieving it as needed. The queueing system serves as an additional buffer to handle bursts of multiple events.

Another thread retrieves events at defined intervals from the queue, creates a copy, and displays it (Figure 1). This sampled data is shown to provide the user with insight into the DAQ process. The software was written this way to ensure that older computers with slower CPU and graphics capabilities can run the program without losing core functionality. The data display and data file format are implemented using ROOT framework [11].

---

* Corresponding author: DmitriyBeznosko@clayton.edu

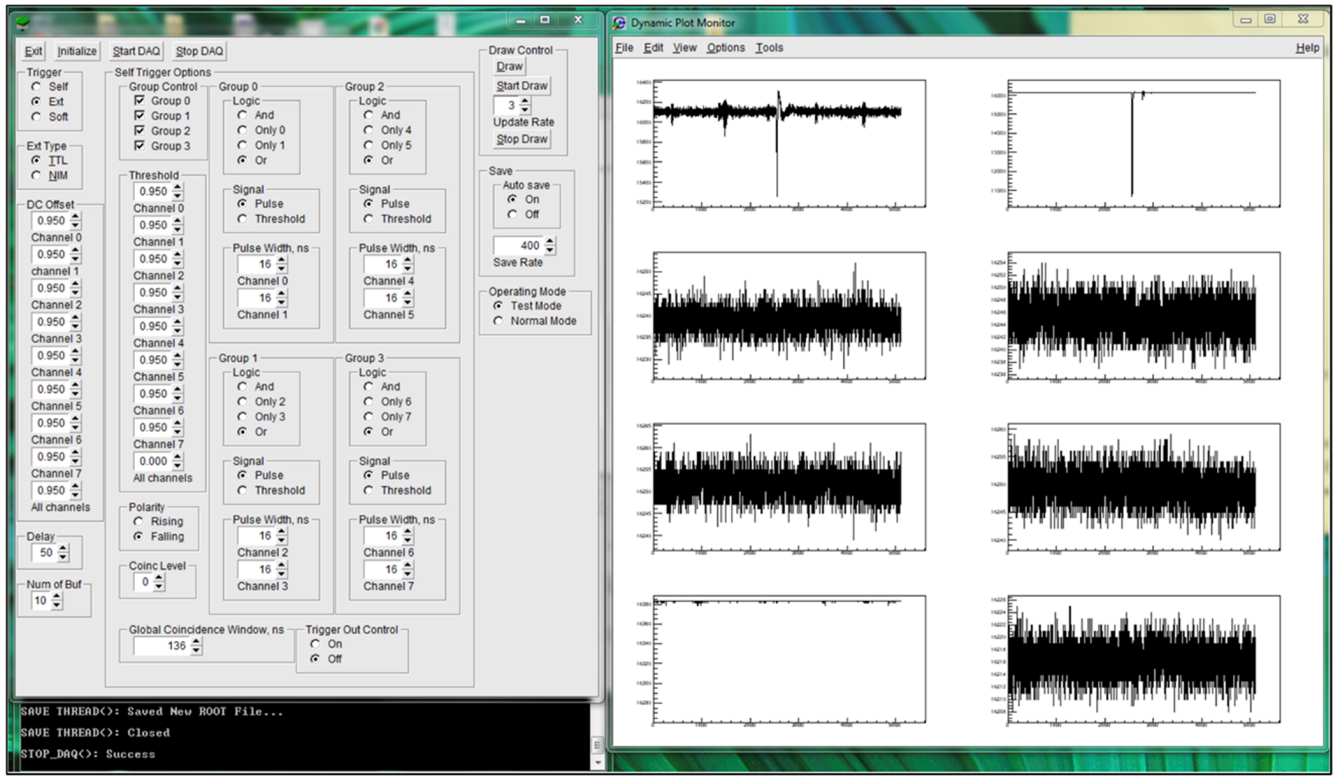

**Fig. 1.** (Right) DAQ control Panel interface. (Left) Data display showing live data from FADC channels 0 - 7.

While the total of 24 channels (3 FADCs with 8 channels on each one) can be handled by the DAQ software, the display only shows the 8 channels from the FADC board that was designated as 'main' during the DAQ initialization.

## 2.1 ROOT Framework

ROOT is a software framework developed by CERN for conducting high-energy physics research. It is widely used for its ability to store, process, visualize, and analyze scientific data efficiently. Working with the ROOT package has several advantages:
1) ROOT provides its file format and data structure as a TTree internal class that is extremely powerful for quick access to large volumes of data - much faster than accessing a text or linear binary file.
2) Data saved in one or more ROOT files can be accessed from a PC, from the Internet and from large-scale file delivery systems. TTree from multiple files with the same data format can be chained together and accessed as a single object that allows you to loop through large amounts of data across those files seamlessly.
3) Powerful mathematical and statistical tools are provided to work with the data.
4) Results can be displayed with bar charts, scatter plots, fitting features and so on. Libraries (with header files) for creating graphical windows and interface elements like control buttons etc. are provided.
5) ROOT provides a set of bindings for easy integration with popular languages, such as Python and R.

### 2.1.1 ROOT File Format:

The DAQ method is implemented using ROOT's built-in TFile class to save the data as .root files. These .root files are binary and are designed to prioritize easy retrieval of information while maintaining a compact file size. They use a hierarchical structure for data organization, with data stored in tree-like structures that help track complex systems. This structure allows for a more dynamic way of accessing relevant data; that is, any event can be accessed without the need to read the entire file sequentially. The data is also compressed during saving to reduce file size.

*2.1.2 TTree Data Structure*

First, each tree is defined, and branches are added to store relevant information. Next, a leaf is defined to represent the actual values stored within a given branch. The value type must be specified, as ROOT recognizes specific types of values. Appendix A lists all available data types.

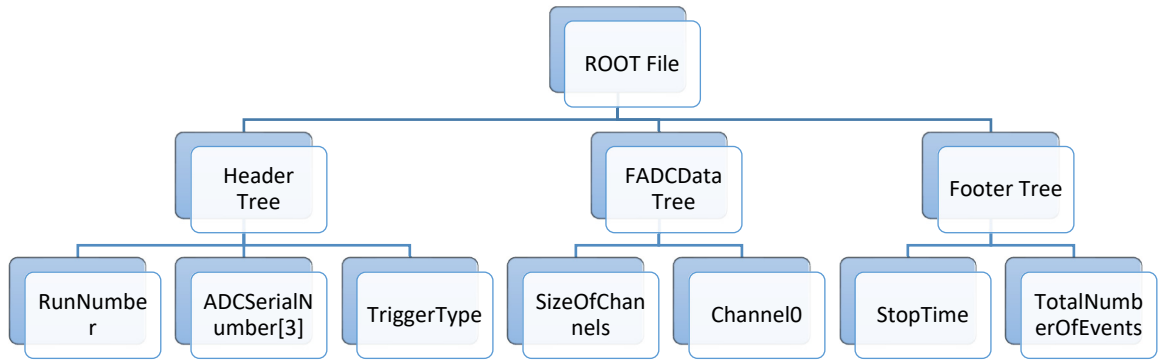

**Fig. 2**: The TTree structure of a data file. The root node defines the structure and branches represent specific parameters.

Consider the tree structure in Figure 2. The defined data root file consists of separate three trees: Header, FADCData, and Footer. Each tree can hold multiple events. In practice, the header and footer typically contain a single event and only the data tree contains multiple events. The data tree name contains the serial number of the FADC board, i.e. FADCData 938.

In Figure 2, under the 'Header Tree', we have a branch called 'TriggerType' that contains a leaf, defined as type 'I', which corresponds to a 32-bit signed integer, corresponding to a single value for the type of trigger used during the data run. The header tree also includes a branch (not shown in Figure 2) called 'GroupLogic' and others as defined in Appendix B. Here is the description of all trees used:

- Header:
  - The header's primary purpose is to define the parameters by which the DAQ was conducted. Parameters include sampling frequency, readout speed, trigger type, trigger polarity, etc.
- Data:
  - The data tree notes the size of a given event and saves FADC data from each individual channel (As Channel0, etc.). The data is written after each trigger, this is done to reduce data loss in the event of malfunction, power loss or software crash.
- Footer:
  - The footer of the file plays a written during closing the file. The footer contains the total number of cosmic events (spanning all previous files in the current run) and time of file closing.

## 3 Conclusions and Future Plans

Currently, the DAQ software is fully functional and capable of obtaining data efficiently. The program architecture was designed with a focus on data acquisition stability and adaptability. The software shows great promise as the default method for FADC control and DAQ functions and will be a foundation for all future software development. The software requires the addition of a new command line interface for a more dynamic design. Development is currently being considered, and preliminary work has been conducted for this functionality.

Limitations: current version lacks the ability to seamlessly adjust data retrieval parameters and switch acquisition modes based on external conditions for optimal data acquisition process. Another limitation is its approach to handling user commands. The current software only allows for the execution of commands through the control panel, as shown in Figure 1.

Command-Line Interface script: the purpose of the future command-line script is to enable the automation of the control of FADC and DAQ within a terminal, while maintaining the ability to execute commands through the control panel manually. The script will be written in Python 3 and will utilize task queuing as the primary method of task execution. It will run a main control method and use Python's subprocess library to manage DAQ software execution.

**Acknowledgements**: This work is supported by NSF LEAPS-MPS Award 2316097

# Appendix A

Below is a list of all valid types that can be used within a TTree structure.

Table 1: List of the parameters in the Footer Tree.

| Reference Character | Type Description |
| --- | --- |
| C | a character string terminated by the O character |
| B | an 8-bit signed integer (Char_t); Treated as a character when in an array |
| b | an 8-bit unsigned integer (UChar_t) |
| S | a 16-bit signed integer (Short_t) |
| s | a 16-bit unsigned integer (UShort_t) |
| I | a 32-bit signed integer (Int_t) |
| i | a 32-bit unsigned integer (UInt_t) |
| F | a 32-bit floating point (Float_t) |
| f | a 21-bit floating point w/ truncated mantissa (Float16_t): 1 for the sign, 8 for the exponent and 23 for the mantissa. |
| D | a 64-bit floating point (Double_t) |
| d | a 32-bit truncated floating point (Double32_t): 1 for the sign, 8 for the exponent and 23 for the mantissa. |
| L | a 64-bit signed integer (Long64_t) |
| l | a 64-bit unsigned integer (ULong_t) |
| G | a long signed integer, stored as 64-bit (Long_t) |
| g | a long unsigned integer, stored as 64-bit (ULong_t) |
| O | [the letter o, not a zero] a Boolean (bool) |

# Appendix B

Below is a list of all specified parameters in the ROOT data file. The ROOT Data Type codes are in Appendix B. Some types are indicated as the standard c++ ones, they will have correspondence to ROOT types depending on your system, typically, int – S and float – F, but this might change depending on the compiler version and other details. Some parameters exist for the data file format compatibility between the projects. Notation '[8]/I' and similar will mean the array of the length as indicated by the number and the type of variables as indicated by the letter, i.e. '[8]/I' is the array with 8 elements, each element being a 32-bit signed integer.

**Table 2:** List of the parameters in the Heater Tree.

| Parameter Name | Data Type | Description |
| --- | --- | --- |
| RunNumber | int | A number for the run. Incremented when the 'Start' is pressed or issued. |
| SamplingFrequency | int | FADC sampling frequency in MHz, fixed at 500 for DT5730, for compatibility. |
| ReadoutSpeed | int | Rate, in Hz, of DAQ software access to FADC buffer, typically 100. |
| ChannelMask | int | Coded trigger channel mask used. |
| TriggerType | int | FADC trigger setting. 0 – software issued, 1 – self, 2 – external trigger. |
| TriggerPolarity | int | Polarity of the external trigger. 1 – TTL (positive polarity), 2 – NIM (negative polarity). |
| FormalDCOffset | [8]/F | The value of the offset of each FADC channel in Volts from 0 ($\pm 1$ for DT5730) set by user. For negative polarity signals, offset should be set to high positive value (about 0.95). |
| RealDCOffset | [8]/F | The value of the offset of each FADC channel as reported by the FADC. Serves for verification. |
| FormalThreshold | [8]/F | The value of the trigger threshold of each FADC channel in Volts from 0 ($\pm 1$ for DT5730) set by user. |
| RealThreshold | [8]/F | The value of the trigger threshold of each FADC channel as reported by the FADC. Serves for verification. |
| StartTime | long long int | A 64-bit field containing the system time in ms of the moment the data acquisition has started. |
| MAXNB | /I | Number of FADCs connected. |
| ADCSerialNumber | [MAXNB]/I | Array of all connected FADC serial numbers. |
| MainBoardSerialNumber | int | Serial number of the FADC that is used for logic to produce the trigger signal. |
| DistanceFromCenter | float | Distance of DAQ unit from detector center position, for format compatibility. |
| TimeDelayCalib | float | Reserved for cable time delays for each channel, in ns. Will be [8]/F in future versions. |
| TriggerDelay | int | Trigger delay in % of the length of the event readout. Sets how much of the data is included before and after the trigger. I.e., calibration data typically has this setting at 50 so that the MIP data is centered in the channel. |
| EventsPerFile | int | Maximum number of events per file as set in DAQ by user. Actual file name have less events. |
| SynchronizationSchema | int | A code for the synch signal used. Set manually if needed, currently is not implemented. |
| NumberOfBuffers | int | The number of buffers allocated per channel by the FADC is $2^{NumberOfBuffers}$. This sets the event size with the smallest being 5110 points (for 1024 buffers), doubling with each step down in value. 1 pint per 2 ns for DT5730. Buffers are used by the FADC to store events before they are read out by the DAQ. |
| GroupLogic | [4]/I | Code for the coincidence logic set for each channel group (4 groups of 2 channels for DT5730). For example, group #0 includes channels 0 and 1. The GroupLogic[0] set to 0 indicates 0 AND 1, set to 1 – only channel 0 is used, set to 2 – only channel 1 is used, set to 3 – 0 OR 1. |
| CoincidenceLevel | int | Coincidence level between the groups – i.e. how many groups need to trigger within the global window to issue a trigger. 1 means that any one group (if active) is needed, x – coincidence of x is required. From 1 to 4. Set by check boxes in the 'Group Control' in DAQ window. |
| SignalTypes | [4]/I | Signal setting of the trigger: 0 – Pulse, 1 – Threshold. Setting is per group of channels. |
| PulseWidth | [8]/I | Logic pulse width for trigger coincidence, in ns, per channel. |
| GlobalWindowWidth | int | Width of the global coincidence window between channel groups. |

**Table 3:** List of the parameters in the FADCData Tree.

| Parameter | Data Type | Description |
|---|---|---|
| SizeOfChannels | /S | Number of points in each channel. Can also be calculated as (5242880/NumberOfBuffers – 10) for the DT5730 FADC. |
| Channel0 … Channel7 | [size]/s | FADC data array of length "SizeOfChannels" and of type s, in ADC bins. 0-7 for 8-channel FADC. |
| TriggerTimeData | /i | A number of counts in the ADC from the start to the trigger of this event. |
| SystemTimeData | /L | A 64-bit field containing the system time in ms of the moment this event was written. |

Table 4: List of the parameters in the FADCData Tree.

| Parameter | Data Type | Description |
|---|---|---|
| TotalNumberOfEvents | /I | Total number of events in the current run, this number can be larger than the number of events in a current file if data run spans multiple files. |
| StopTime | /L | A 64-bit field containing the system time in ms of the moment the run was stopped, or the file was closed. |